\documentclass[prb,reprint,a4paper,showpacs,citeautoscript,floatfix]{revtex4-1}

\pdfoutput=1
\usepackage[charter]{mathdesign}
\usepackage{amsmath}

\usepackage[pdftex]{graphicx}
\usepackage{microtype}
\usepackage{bm}\let\vec\bm

\usepackage[svgnames]{xcolor}
\usepackage[
	colorlinks=True,linkcolor=DarkRed,citecolor=ForestGreen,urlcolor=MediumBlue,
	pdfstartview=FitH,bookmarks=False,pdfpagemode=UseNone
]{hyperref}

\begin{document}

\title{Tunneling conductance and local density of states in tight-binding junctions}

\author{C. Berthod}
\author{T. Giamarchi}
\affiliation{DPMC-MaNEP, Universit{\'e} de Gen{\`e}ve, 24 quai Ernest-Ansermet, 1211 Gen{\`e}ve 4, Switzerland}

\date{October 12, 2011}

\begin{abstract}

We study the relationship between the differential conductance and the local density of states in tight-binding tunnel junctions where the junction' geometry can be varied between the point-contact and the planar-contact limits. The conductances are found to differ significantly in these two limiting cases. We also examine how the matrix element influences the tunneling characteristics and produces contrast in a simple model of scanning tunneling microscope (STM). Some implications regarding the interpretation of STM spectroscopic data in the cuprates are discussed. The calculations are carried out within the real-space Keldysh formalism.

\end{abstract}

\maketitle

\section{Introduction}

The spectrum of one-particle excitations is a rich source of information in the study of condensed-matter systems \cite{Mahan-2000}. Among the various tools available to measure the low-energy excitations, tunneling has become a prominent technique for electronic materials, due to its high energy resolution and the possibility to set up local probes \cite{Wolf-1985}. In a tunneling experiment, electrons are extracted from or injected into a sample through a classically forbidden barrier, thereby probing the spectrum of excitations below and above the Fermi energy. The practical realization requires to keep the sample and the probing electrode sufficiently isolated that they do not significantly interact, and sufficiently close that a measurable number of electrons  can tunnel from one to the other. This has been achieved in a variety of ways, including the planar junction with a thin insulating layer \cite{Sze-1981} and the vacuum junctions obtained, e.g., by breaking the sample \cite{Moreland-1985, Agrait-2003} or using a scanning tunneling microscope (STM) tip \cite{Wiesendanger-1993, Briggs-1999, Fischer-2007}. The precise relationship between the current-voltage characteristics measured at the tunnel junction and the spectrum of excitations has been a matter of continuous research since half a century \cite{Harrison-1961, Bardeen-1961, *Bardeen-1962, Cohen-1962, Prange-1963, Caroli-1971a, *Caroli-1971b, *Caroli-1972, Feuchtwang-1974a, *Feuchtwang-1974b, *Feuchtwang-1975, *Feuchtwang-1976, Tersoff-1983, *Tersoff-1985, *Tersoff-1989, Chen-1988, *Chen-1990a, *Chen-1990b, Ferrer-1988, Lucas-1988, Noguera-1990, *Sacks-1991, Pendry-1991, Meir-1992, Todorov-1993, Frederiksen-2007, Passoni-2007, Ryndyk-2009}.

The problem may be divided into three distinct questions: (i) is the tunneling-Hamiltonian formalism \cite{Cohen-1962} appropriate, and if yes, how should the tunneling Hamiltonian be defined? More elaborate approaches have suggested that the tunneling Hamiltonian can be a good approximation \cite{Prange-1963, Noguera-1990, Pendry-1991}, but the proper definition of the tunneling matrix element remains a difficult problem lacking a general solution, that would remain valid for interacting systems and out-of-equilibrium conditions. (ii) Provided that the tunneling-Hamiltonian gives a good description of the system and that the tunneling matrix elements are known, how does the current relate to the excitation spectrum in the electrodes? In particular, is it possible to measure the density of states (DOS) or the local DOS (LDOS) when using local probes? This question is crucial for the interpretation of experimental data, but has not been thoroughly investigated. (iii) Although surfaces are of great interest by themselves, tunneling spectroscopy is often conducted with the aim of addressing bulk properties. Therefore, when it comes to the interpretation of tunneling spectra, the question invariably arises, whether the measurements performed by connecting two surfaces are representative of the bulk materials. We will not address (i) and (iii), but turn our attention to (ii), which is a well-posed theoretical problem. In order to cope with (i), we start right away from a tight-binding Hamiltonian containing two electrodes and a tunneling term. In order to avoid (iii), we shall study systems in which the distinction between the bulk and the surface is irrelevant, at least for the electrode representing the sample.

The goal of the present study is to explore, using minimal models, how the junction' geometry and the specifics of the tunneling term influence the current-voltage characteristics $I(V)$. We are particularly interested in comparing the differential conductance $\sigma(V)=dI/dV$ and the DOS, since these two quantities are commonly believed to be roughly proportional to each other. The DOS and, more generally, the one-electron excitations, are very conveniently described by means of the Green's function, which is an energy-dependent nonlocal quantity $G(\vec{x},\vec{x}',\omega)$ in the real-space representation \cite{Mahan-2000}. The LDOS, in particular, is directly related to the diagonal elements of $G$. The real-space formulation is best suited for our purposes, as we shall investigate geometrical effects and consider systems that break translation invariance. The exact current $I(V)$ can be expressed in terms of the real-space Green's functions using the Keldysh nonequilibrium formalism \cite{Rammer-1986}. In Sec.~\ref{sec:model}, we recall this formalism, which leads to a compact formula for the current [Eqs.~(\ref{eq:current}) and (\ref{eq:GK})] involving only the Green's functions of the electrodes and the tunneling amplitudes \cite{Caroli-1971a, Ferrer-1988}. In Secs.~\ref{sec:T} and \ref{sec:3D}, we use this formula to compute the $I(V)$ characteristics of two simple tight-binding models designed in order to highlight the role of geometry and dimensionality on the relationship between $\sigma(V)$ and the DOS. These models show, in particular, that Harrison's cancellation argument\cite{Harrison-1961}, which explains the ohmic behavior of planar junctions between simple metals, has a relatively narrow range of validity. In Sec.~\ref{sec:STM}, we introduce a toy model of STM and discuss the role of the matrix element in the imaging process and the local spectroscopy. Finally, our main results and their implications are summarized and discussed in Sec.~\ref{sec:discussion}.

\section{Model and formalism}
\label{sec:model}

The hallmark of vacuum tunnel junctions, when compared to other electrical contacts, is that the two electrodes remain electrically well isolated when the junction is formed. This is due to the presence of a high barrier strongly reducing the wave-function overlap. In this situation, a good starting point is to represent the junction by a tunnel Hamiltonian that describes single-electron transfers across the barrier \cite{Cohen-1962}. The electrodes, on the other hand, are assumed to be in thermal equilibrium, and the Hamiltonian describing them to remain unchanged in the presence of the junction, apart from the electrical bias applied to one of the contacts. This assumption can only be valid if the potential drop of the biased junction takes place entirely in the region of the barrier. The extent to which these various assumptions apply to realistic experimental setups is obviously a difficult one, and we do not aim to address it here. We consider instead a class of ideal tight-binding junctions where these assumptions are build in from the start. The junction's differential conductance can then be evaluated exactly without further approximation, and the relation between the conductance and the LDOS can be inspected. In this section, we first review the general formula for the conductance in terms of the Green's functions of the two electrodes and the tunneling amplitudes, and then consider a few remarkable special situations where the expression of the conductance can be simplified further. Although we restrict our applications to the case of noninteracting single-band metals, many of the quoted expressions apply equally to multiband interacting systems, as long as the correlations \emph{induced by the tunneling term} are negligible in comparison to the correlations intrinsic to the electrodes.

\subsection{General formula for the differential conductance}

Consider two electrodes, left ($L$) and right ($R$), described by the lattice tight-binding Hamiltonians $\mathcal{H}_L$ and $\mathcal{H}_R$, respectively. The electrodes being initially decoupled means that these Hamiltonians commute: $[\mathcal{H}_L,\mathcal{H}_R]=0$. We denote the lattice sites in $L$ and $R$ by $\vec{l}$ and $\vec{r}$, and the corresponding electron field operators by $\psi^{\dagger}(\vec{l})$ and $\psi^{\dagger}(\vec{r})$. The spin coordinate of the electrons is omitted throughout. The two electrodes are coupled by a tunneling term
	\begin{equation}\label{eq:tunneling-Hamiltonian}
		\mathcal{H}_T=\sum_{\vec{l}\vec{r}}T(\vec{l},\vec{r})\psi^{\dagger}(\vec{l})
		\psi(\vec{r})+\text{H.c.},
	\end{equation}
so that the system's Hamiltonian is $\mathcal{H}=\mathcal{H}_L+\mathcal{H}_R+\mathcal{H}_T$. The tunneling amplitudes $T(\vec{l},\vec{r})$ are symmetric, and we also assume that they are real. When a bias is applied to $L$, a steady-state current $I$ is established. The calculation of the current is outlined in Appendix~\ref{sec:app1} and gives
	\begin{equation}\label{eq:current}
		I=\frac{e}{h}\int d\omega\,\text{Re}\,\text{Tr}\,TG^K(\omega).
	\end{equation}
$T$ is the matrix with elements $T(\vec{l},\vec{r})$, and $G^K$ is the Keldysh matrix Green's function defined in the time domain by $G^K(\vec{r},\vec{l},t)=-i\langle[\psi(\vec{r},t),\psi^{\dagger}(\vec{l},0)]\rangle$ with $\langle\cdots\rangle$ the thermodynamic average with respect to $\mathcal{H}$, the fully formed junction. The trace is over the sites $\vec{l}$ of $L$. Hence the current sums all closed paths from a point $\vec{l}$ to itself with excursion in $R$ through the barrier. For convenience, it is desirable to express $G^K(\omega)$ in terms of the Green's functions of the \emph{isolated} electrodes. At this step, we must implement the assumptions that the bias-induced potential drop takes place in the barrier region, and that the two electrodes are in thermal equilibrium at the chemical potentials $\mu_L$ and $\mu_R$. This gives (see Appendix~\ref{sec:app1})
	\begin{multline}\label{eq:GK}
		TG^K=(\openone-TG_R^+TG_L^+)^{-1}\Big[(1-2f_L)TG_R^+T(G_L^+-G_L^-)\\
		+(1-2f_R)T(G_R^+-G_R^-)TG_L^-\Big](\openone-TG_R^-TG_L^-)^{-1}.
	\end{multline}
$G_{L,R}^{\pm}(\omega)$ are the retarded ($+$) and advanced ($-$) matrix Green's functions in $L$ and $R$. In the time domain, they read $G_L^{\pm}(\vec{l}_1,\vec{l}_2,t)=\mp i\theta(\pm t)\langle[\psi(\vec{l}_1,t),\psi^{\dagger}(\vec{l}_2,0)]_+\rangle$ with $[\,\cdot\,,\cdot\,]_+$ being the anticommutator, and, similarly, for $G_R^{\pm}$. In the frequency domain, we have $G_{L,R}^-(\omega)=[G_{L,R}^+(\omega)]^{\dagger}$. $f_{L,R}(\omega)$ is the Fermi distribution measured from $\mu_{L,R}$. Equations~(\ref{eq:current}) and (\ref{eq:GK}) allow to evaluate the current-voltage characteristics $I(V)$ as well as the tunneling conductance $dI/dV$ once the electrodes and the tunneling amplitudes are specified. The size of the matrices to be considered is given by the number of sites connected by nonvanishing tunneling matrix elements. The propagators $G_{L,R}^{\pm}(\omega)$ must be calculated for the semiinfinite electrodes, but they are only needed among this same subset of sites for calculating the current. It should be noted that while Eq.~(\ref{eq:current}) is general, Eq.~(\ref{eq:GK}) is not exact if there are interactions in the electrodes. For noninteracting electrodes, Eqs.~(\ref{eq:current}) and (\ref{eq:GK}) reduce to Eq.~(37) of Ref.~\onlinecite{Todorov-1993} (see Appendix~\ref{sec:app2}). The derivation of Ref.~\onlinecite{Todorov-1993} explicitly relies on the assumption of independent electrons, and seems difficult to generalize. An advantage of the Keldysh formulation is that interactions can be easily included, at least formally. In the case of interacting electrodes, it can be, for example, shown that the current obtained by inserting the \emph{interacting} Green's functions in Eq.~(\ref{eq:GK}) correctly accounts for the correlations intrinsic to the electrodes, but disregards the correlations induced by the tunneling term. Within their domain of validity, Eqs.~(\ref{eq:current}) and (\ref{eq:GK}) are exact at all orders in $T(\vec{l},\vec{r})$ and $V$ and for arbitrary temperatures.

\begin{figure*}[t!]
\includegraphics[width=1.7\columnwidth]{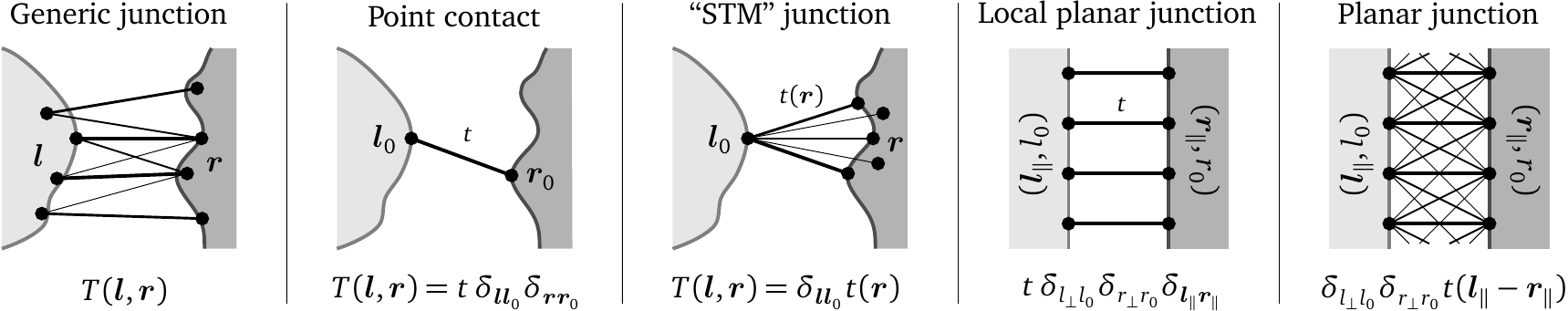}
\caption{\label{fig:fig1}
Generic tight-binding tunnel junction (left), and particular examples with the corresponding tunneling matrix element. The black dots represent discrete points of an arbitrary tight-binding lattice.
}
\end{figure*}

In vacuum tunneling experiments, one generally uses a probe to measure a sample. From now on, we shall consider $L$ as the probe and $R$ as the sample. It is convenient to measure energies from the chemical potential of the sample: we thus take $\mu_R\equiv 0$ and $\mu_L=eV$. As a result, $f_R(\omega)=[\exp(\beta\omega)+1]^{-1}\equiv f(\omega)$ with $\beta$ the inverse temperature, and $f_L(\omega)=f(\omega-eV)$. With this convention, $G_R^{\pm}(\omega)$ to be used in Eq.~(\ref{eq:GK}) is the same as that for the isolated sample $R$, while $G_L^{\pm}(\omega)$ should be the Green's function in $L$ with the origin of energies shifted by $eV$, i.e., $G_L^{\pm}(\omega)=G_{L,0}^{\pm}(\omega-eV)$ with $G_{L,0}^{\pm}$ the Green's function in the absence of bias. Therefore two kinds of terms appear when taking the bias derivative $dI/dV$: those from the bias dependence of $f_L$, and those from the bias dependence of $G_L^{\pm}$. For $L$ to be a useful probe, its properties must vary slowly with energy, so that the latter terms are expected to be small compared to the former. We consequently split the differential conductance and define
	\begin{equation}\label{eq:dIdV}
		\sigma(V)=\frac{dI}{dV}=\tilde{\sigma}(V)+\delta\sigma(V).
	\end{equation}
Retaining only the bias dependence of $f_L$ in Eq.~(\ref{eq:GK}), the dominant contribution to the conductance is
	\begin{multline}\label{eq:sigmatilde}
		\tilde{\sigma}(V)=\frac{e^2}{h}\int d\omega\,[-f'(\omega-eV)]2\text{Re}\,\text{Tr}\,
		TG_R^+T(G_L^--G_L^+)\\\times(\openone-TG_R^-TG_L^-)^{-1}(\openone-TG_R^+TG_L^+)^{-1}.
	\end{multline}
The $\omega$ dependence of $G_{L,R}^{\pm}$ is implicit, $f'$ is the derivative of $f$, and we have used the cyclic property of the trace. Equation~(\ref{eq:sigmatilde}) is a convenient starting point for performing analytical calculations: at zero temperature, $-f'(\omega-eV)$ becomes $\delta(\omega-eV)$, and the energy integral drops. $\tilde{\sigma}(V)$ then only depends on $G_L^{\pm}(eV)=G_{L,0}^{\pm}(0)$ and $G_R^{\pm}(eV)$. It is also seen that the main effect of a finite temperature on $\tilde{\sigma}(V)$, apart from small contributions related to the bias dependence of $G_L^{\pm}$, is to thermally broaden the zero-temperature curve. We have checked on simple models that $\tilde{\sigma}(V)$ gives the same conductance as the Landauer-B{\"u}ttiker formula \cite{Buttiker-1986} at zero temperature, provided that the bias $eV$ in $L$ is taken into account when calculating the scattering matrix. Hence $\delta\sigma(V)$ corresponds to the difference between the differential and the ballistic conductances.

\subsection{Particular cases}

In this section, we present more explicit expressions for the conductance, in a few limiting situations depicted in Fig.~\ref{fig:fig1}. While for a generic junction the matrix structure in Eq.~(\ref{eq:sigmatilde}) cannot be simplified further, in these cases the conductance can be reduced to a scalar form. The first case is the ``point contact'', by which we mean a junction such that the tunneling is only possible from a single site $\vec{l}_0$ in $L$ to a single site $\vec{r}_0$ in $R$. The tunneling amplitude is $T(\vec{l},\vec{r})=t\delta_{\vec{l}\vec{l}_0}\delta_{\vec{r}\vec{r}_0}$. In reciprocal space, the matrix element is therefore independent of momentum, $|T_{\vec{k}\vec{k}'}|^2\equiv t^2$. The latter assumption is quite common---although often implicit---in theoretical studies of tunnel junctions. The fact that it is a drastic simplification appears somewhat more clearly when it is formulated in real space. The locality of the tunneling amplitude implies that all matrices in Eq.~(\ref{eq:sigmatilde}) become scalars. Using the definition of the LDOS, e.g., $N_L(\vec{l}_0,\omega)=-\frac{1}{\pi}\text{Im}\,G_L^+(\vec{l}_0,\vec{l}_0,\omega)=\frac{1}{2\pi i}[G_L^-(\vec{l}_0,\vec{l}_0,\omega)-G_L^+(\vec{l}_0,\vec{l}_0,\omega)]$, one readily obtains \cite{Ferrer-1988}
	\begin{multline}\label{eq:point_contact}
		\tilde{\sigma}(V)=\frac{e^2}{h}\int d\omega\,[-f'(\omega-eV)]N_L(\vec{l}_0,\omega)\\
		\times\frac{(2\pi t)^2N_R(\vec{r}_0,\omega)}
		{|1-t^2G_L^+(\vec{l}_0,\vec{l}_0,\omega)G_R^+(\vec{r}_0,\vec{r}_0,\omega)|^2}\\[0.5em]
		\qquad\text{(point contact)}.
	\end{multline}
At zero temperature and in the weak tunneling regime $t\to 0$, we thus have an exact proportionality between $\tilde{\sigma}(V)$ and the sample LDOS at the point $\vec{r}_0$: $\tilde{\sigma}(V)\propto N_{L,0}(\vec{l}_0,0)N_R(\vec{r}_0,eV)+\mathcal{O}(t^4)$. $N_{L,0}(\vec{l}_0,0)\equiv N_L(\vec{l}_0,eV)$ is the LDOS at the Fermi energy and at the point $\vec{l}_0$ in $L$. This proportionality is lost as $t$ increases and becomes comparable to the typical energy scales in $L$ and $R$, so that the product $t^2G_L^+G_R^+$ in the denominator becomes of order unity. In the class of junctions that we are considering, the weak point-contact limit is the only case where there can be a strict proportionality between the tunneling conductance and the LDOS \cite{Gramespacher-1997, *Gramespacher-1998, *Gramespacher-1999}.

A second case of interest is when the tunneling is possible from a single point $\vec{l}_0$ in $L$ to several points $\vec{r}$ in $R$. This can be considered as a crude model for an STM with some dispersion of the electrons tunneling from the tip. The matrix element reads $T(\vec{l},\vec{r})=\delta_{\vec{l}\vec{l}_0}t(\vec{r})$. It is still possible to invert the matrices analytically in this case, and one obtains
	\begin{multline}\label{eq:STM_junction}
		\tilde{\sigma}(V)=\frac{e^2}{h}\int d\omega\,[-f'(\omega-eV)]N_L(\vec{l}_0,\omega)\\
		\times\frac{(2\pi)^2\displaystyle\sum_{\vec{r}_1\vec{r}_2}
		t(\vec{r}_1){\textstyle\left(-\frac{1}{\pi}\right)}
		\text{Im}[G_R^+(\vec{r}_1,\vec{r}_2,\omega)]t(\vec{r}_2)}{\displaystyle
		\Big|1-G_L^+(\vec{l}_0,\vec{l}_0,\omega)\sum_{\vec{r}_1\vec{r}_2}
		t(\vec{r}_1)G_R^+(\vec{r}_1,\vec{r}_2,\omega)t(\vec{r}_2)\Big|^2}\\[0.5em]
		\qquad\text{(``STM'' junction)}.
	\end{multline}
When compared to Eq.~(\ref{eq:point_contact})---that is recovered by taking $t(\vec{r})=t\delta_{\vec{r}\vec{r}_0}$---Eq.~(\ref{eq:STM_junction}) includes all excursions within $R$ among sites that are connected back to $\vec{l}_0$. This leads to interference between the various paths and to the loss of a simple proportionality between $\tilde{\sigma}(V)$ and the LDOS in $R$.

The planar analog of the point contact is a junction such that tunneling is only possible between sites facing each other on the two surfaces. The matrix element is $T(\vec{l},\vec{r})=t\delta_{l_{\perp}l_0}\delta_{r_{\perp}r_0}\delta_{\vec{l}_{\parallel}\vec{r}_{\parallel}}$, where $l_{\perp}$ and $r_{\perp}$ denote the spatial coordinates in the direction perpendicular to the junction's plane, $l_0$ and $r_0$ are the coordinates of the surfaces of $L$ and $R$, and $\vec{l}_{\parallel}$ and $\vec{r}_{\parallel}$ denote the coordinates in the plane of the junction. In the momentum representation $\vec{k}_{\parallel}$, all matrices in Eq.~(\ref{eq:sigmatilde}) again become scalars, leading to
	\begin{multline}\label{eq:local_planar_junction}
		\tilde{\sigma}(V)=\frac{e^2}{h}\int d\omega\,[-f'(\omega-eV)]\sum_{\vec{k}_{\parallel}}
		A_L(\vec{k}_{\parallel},l_0,l_0,\omega)\\
		\times\frac{(2\pi t)^2A_R(\vec{k}_{\parallel},r_0,r_0,\omega)}
		{|1-t^2G_L^+(\vec{k}_{\parallel},l_0,l_0,\omega)
		G_R^+(\vec{k}_{\parallel},r_0,r_0,\omega)|^2}\\[0.5em]
		\qquad\text{(local planar junction)}.
	\end{multline}
We have introduced the local spectral functions $A_L(\vec{k}_{\parallel},l_1,l_2,\omega)=\frac{1}{2\pi i}[G_L^-(\vec{k}_{\parallel},l_1,l_2,\omega)-G_L^+(\vec{k}_{\parallel},l_1,l_2,\omega)]$, etc., whose diagonal part $A_L(\vec{k}_{\parallel},l_0,l_0,\omega)$ has the physical meaning of a $\vec{k}_{\parallel}$-resolved LDOS on the surface of $L$. The crucial difference between Eqs.~(\ref{eq:point_contact}) and (\ref{eq:local_planar_junction}) is the conservation of in-plane momentum in the latter, which imposes to match states with the same $\vec{k}_{\parallel}$ across the barrier. This restriction implies that the conductance per channel must be smaller in the planar than in the point-contact case, as we shall see below. Finally, we may consider a straightforward generalization of the local planar junction by including additional tunneling paths as depicted in the rightmost panel of Fig.~\ref{fig:fig1}. The matrix element becomes $T(\vec{l},\vec{r})=\delta_{l_{\perp}l_0}\delta_{r_{\perp}r_0}t(\vec{l}_{\parallel}-\vec{r}_{\parallel})$, but since the translation invariance is preserved the structure of the matrices remains the same as in the previous situation:
	\begin{multline}\label{eq:planar_junction}
		\tilde{\sigma}(V)=\frac{e^2}{h}\int d\omega\,[-f'(\omega-eV)]\sum_{\vec{k}_{\parallel}}
		A_L(\vec{k}_{\parallel},l_0,l_0,\omega)\\
		\times\frac{[2\pi t(\vec{k}_{\parallel})]^2A_R(\vec{k}_{\parallel},r_0,r_0,\omega)}
		{|1-t^2(\vec{k}_{\parallel})G_L^+(\vec{k}_{\parallel},
		l_0,l_0,\omega)G_R^+(\vec{k}_{\parallel},r_0,r_0,\omega)|^2}\\[0.5em]
		\qquad\text{(Planar junction)},
	\end{multline}
with $t(\vec{k}_{\parallel})$ the Fourier transform of $t(\vec{l}_{\parallel}-\vec{r}_{\parallel})$.

In the subsequent sections, we shall use Eqs.~(\ref{eq:current})--(\ref{eq:dIdV}) to compute numerically the exact conductance, including the small correction $\delta\sigma$ due to the electronic structure in $L$. For the discussion and interpretation of the results, however, extensive use will be made of the simpler expressions (\ref{eq:sigmatilde})--(\ref{eq:planar_junction}).

\section{Two-dimensional T-shaped junction}
\label{sec:T}

\begin{figure}[t]
\includegraphics[width=0.45\columnwidth]{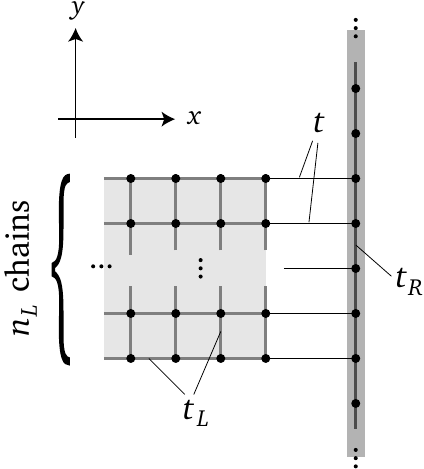}
\caption{\label{fig:fig2}
Tunnel junction between a semiinfinite rod and an infinite one-dimensional chain. For $n_L=1$, the junction is a T-shaped point contact while for $n_L=\infty$, it is a local planar junction according to the terminology of Fig.~\ref{fig:fig1}.
}
\end{figure}

We have seen that in the point-contact limit and in the weak-tunneling regime the conductance is proportional to the sample LDOS. This proportionality is lost when the coupling increases. It is also expected to be lost in the planar geometry due to the conservation of in-plane momentum. In order to investigate these various limits and the transition between them, we consider the system represented in Fig.~\ref{fig:fig2}. The probe $L$ is made of $n_L$ interconnected semiinfinite chains forming a flat rod. The Hamiltonian $\mathcal{H}_L$ involves only nearest-neighbor hopping with energy $t_L$. The sample $R$ is an infinite linear chain with nearest-neighbor hopping $t_R$. We consider $L$ and $R$ to be half-filled for simplicity. Finally, the tunneling term $\mathcal{H}_T$ connects neighboring sites at the interface with the hopping $t$. For $n_L=1$, we have a point contact, while for $n_L=\infty$, we have a local planar junction according to the terminology of Fig.~\ref{fig:fig1}. This is perhaps the simplest model to address the relation between the conductance and the LDOS. The sample has a position-independent LDOS given by
	\begin{equation}\label{eq:T-NR}
		N_R(\omega)=\frac{1}{2\pi|t_R|}\text{Re}\left\{\frac{1}
		{\sqrt{1-\left[\omega/(2t_R)\right]^2}}\right\}.
	\end{equation}
This LDOS has square-root band edges divergences at $\pm2|t_R|$, that are easily identified if they appear in the differential conductance. We need the retarded and advanced Green's functions in both electrodes. In $R$, they can be given in closed form:
	\begin{eqnarray}\label{eq:T-GR}
		\nonumber
		G_R^{\pm}(r_1,r_2,\omega)&=&\int_{-\pi}^{\pi}\frac{dk}{2\pi}
		\frac{e^{ik(r_1-r_2)}}{\omega-2t_R\cos k\pm i0^+}\\
		&=&\frac{1}{2t_R}R_{r_1-r_2}\left(\frac{\omega\pm i0^+}{2t_R}\right),
\end{eqnarray}
\begin{eqnarray}
		\nonumber
		R_n(z)&=&\frac{1}{\sqrt{z^2-1}}\left[\theta(\text{Re}\,z)\left(z-\sqrt{z^2-1}\right)^{|n|}
		\right. \\
		\nonumber
		&&\hspace*{1.3cm}\left.-\theta(-\text{Re}\,z)\left(z+\sqrt{z^2-1}\right)^{|n|}\right].
	\end{eqnarray}
On the surface of $L$ ($x=0$), we may express the Green's function in terms of the single-electron wave functions $\varphi_{kq}(x,y)$ as
	\begin{equation}
		G_L^{\pm}(l_1,l_2,\omega)=\sum_q\int_{-\pi}^{\pi}\frac{dk}{2\pi}\,\frac{\varphi_{kq}(0,l_1)
		\varphi_{kq}^*(0,l_2)}{\omega-eV-\varepsilon_{kq}\pm i0^+}.
	\end{equation}
Indices $l_1, l_2=1,\ldots,n_L$ run on the chains in the $y$ direction, $\varepsilon_{kq}$ are the single-electron energies with $k$ ($q$) the momentum along $x$ ($y$), and we have introduced the bias $eV$ as discussed in the previous section. For open boundary conditions, $\varphi_{kq}(x,0)=\varphi_{kq}(x,n_L+1)=\varphi_{kq}(1,l)=0$, the wave functions that solve $\mathcal{H}_L$ are
	\begin{equation}
		\varphi_{kq}(x,l)=\left[e^{ik(x-1)}-e^{ik(1-x)}\right]\sqrt{\frac{1}{n_L+1}}\sin(ql).
	\end{equation}
The energies are $\varepsilon_{kq}=2t_L(\cos k+\cos q)$ and the discrete $q$ momenta are $q=\frac{\lambda\pi}{n_L+1}$, $\lambda=1,\ldots,n_L$. With this, the $k$ integral can be evaluated and the Green's function becomes
	\begin{eqnarray}\label{eq:T-GL}
		\nonumber
		G_L^{\pm}(l_1,l_2,\omega)&=&\frac{2/t_L}{n_L+1}\sum_q\sin(ql_1)\sin(ql_2)\\
		&&\hspace{0.2cm}\times L\left(\frac{\omega-eV\pm i0^+}{2t_L}-\cos q\right)\\
		\nonumber
		L(z)&=&z-\sqrt{z+1}\sqrt{z-1}.
	\end{eqnarray}
The matrices $G_{L,R}^{\pm}$ of Eqs.~(\ref{eq:T-GR}) and (\ref{eq:T-GL}), together with the matrix $T$, which is just $t\times\openone$ in this case, must be substituted into Eqs.~(\ref{eq:current})--(\ref{eq:dIdV}) for evaluating the differential conductance.

\begin{figure}[t]
\includegraphics[width=\columnwidth]{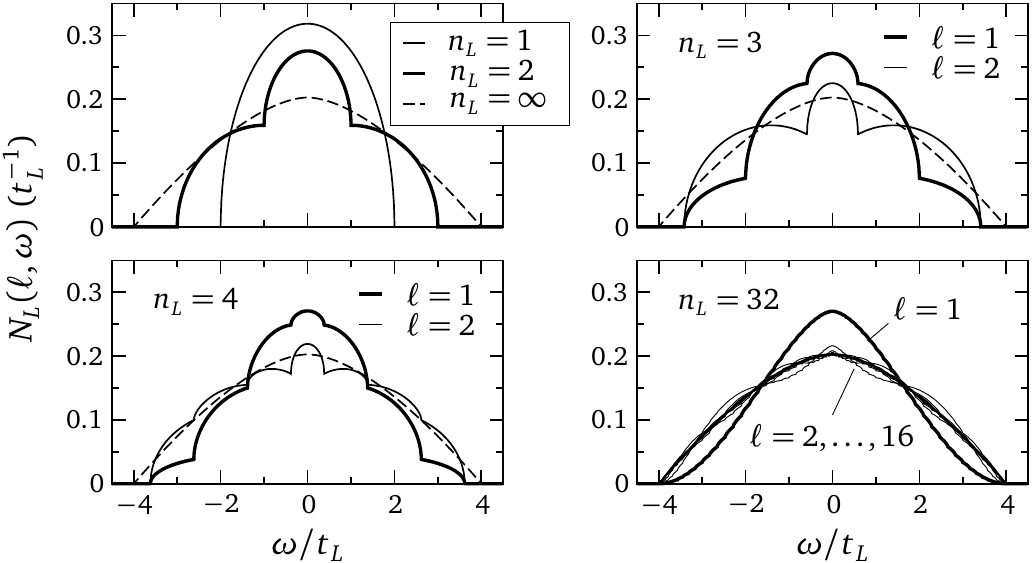}
\caption{\label{fig:fig3}
Local density of states on the surface of the left system in Fig.~\ref{fig:fig2}. As $n_L$ increases, the LDOS in the central portion of the surface approaches the limiting value for $n_L=\infty$, while the LDOS close to $l=1$ has a different shape.
}
\end{figure}

The LDOS at the surface of $L$ can be obtained from $G_L^+$ as $N_L(l,\omega)=(-1/\pi)\text{Im}\,G_L^+(l,l,\omega)$; it is neither energy nor position independent due to the open boundaries. Since we use $L$ as a probe, some understanding of $N_L(l,\omega)$ is necessary before discussing the conductance. In Fig.~\ref{fig:fig3}, we display $N_L(l,\omega)$ for various values of $n_L$. At $n_L=1$, there is only one point in contact with the sample $R$. The LDOS at that point is smooth and takes the form of a half-circle,
	\begin{equation}
		N_L^{(n_L=1)}(\omega)=\frac{1}{\pi|t_L|}\text{Re}\,
		\sqrt{1-\left[\omega/(2t_L)\right]^2},
	\end{equation}
with a half-bandwidth of $2|t_L|$. This contrasts with the LDOS far in the chain ($x\ll0$), which develops band-edge square-root divergences and approaches the form given by Eq.~(\ref{eq:T-NR}) with $t_L$ substituted for $t_R$. The absence of divergences in $N_L(l,\omega)$ qualifies $L$ as a convenient probe. This remains true for $n_L>1$, although additional structures develop in the LDOS. For $n_L=2$, the two sites on the surface have the same LDOS by inversion symmetry, with cusps at $\omega=\pm|t_L|$, and the half-bandwidth increases to $3|t_L|$. For $n_L=3$, the LDOS on the central site ($l=2$) and on the borders ($l=1$, $l=3$) have the same bandwidth but different energy dependencies, and the same applies to all $n_L\geqslant 3$. The general trend is that the LDOS becomes smoother and smoother as $n_L$ increases, and approaches the limiting value for $n_L=\infty$ with a half-bandwidth of $4|t_L|$, except at the borders $l=1$ and $l=n_L$ where a different limiting behavior is found. With increasing $n_L$, the characteristic energy of the structures in the LDOS decreases, but at the same time their amplitude is reduced. For $L$ to be a good probe at any $n_L$, the structures in $N_L(l,\omega)$ should not prevent the identification in the conductance of structures due to the sample $R$. This can be achieved by taking $t_L\gg t_R$.

\begin{figure}[tb]
\includegraphics[width=0.9\columnwidth]{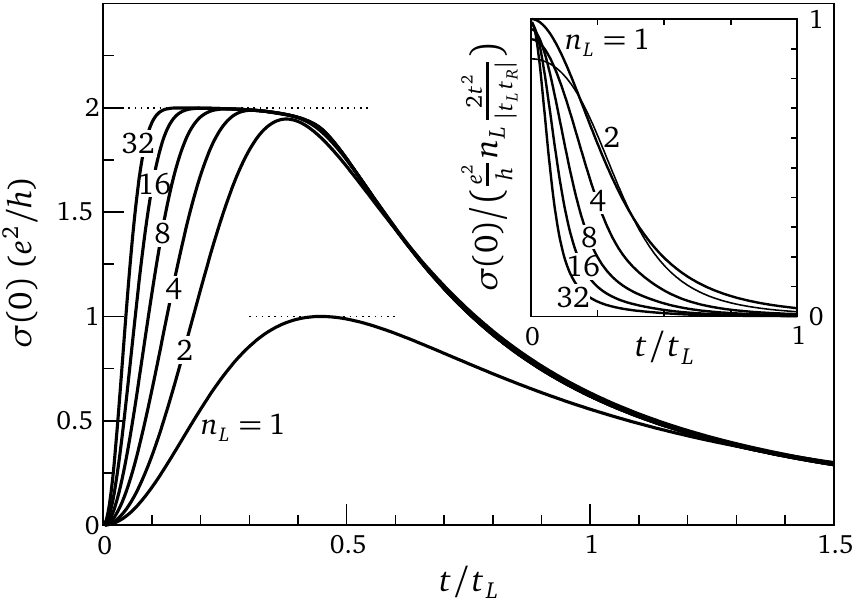}
\caption{\label{fig:fig4}
Zero-bias differential conductance of the T-shaped junction depicted in Fig.~\ref{fig:fig2} for $t_L=10t_R$ and zero temperature. The inset shows the same data rescaled as indicated.
}
\end{figure}

We have calculated numerically the conductance of the T-shaped junction of Fig.~\ref{fig:fig2} as a function of $n_L$, $t$, and $V$ for $t_L=10t_R$ and zero temperature. The zero-bias conductance is presented in Fig.~\ref{fig:fig4}. In the tunneling regime where $t\ll t_L,t_R$, $\sigma(0)$ increases as $t^2$ and is roughly proportional to the contact size $n_L$ as shown in the inset. Specifically, we have
	\begin{equation}
		\sigma(0)=\frac{e^2}{h}a(n_L)n_L\frac{2t^2}{|t_Lt_R|}+\mathcal{O}(t^4)
	\end{equation}
with $a(n_L)$ a slow function of $n_L$: $a(1)=1$, $a(2)=\sqrt{3}/2$, and $a(n_L\gg1)\to1$. $a(2)$ is equal to the ratio of the zero-energy LDOS in $L$ for $n_L=2$ and $n_L=1$. Hence the reduced conductance for $n_L=2$ reflects the reduced zero-energy LDOS seen in Fig.~\ref{fig:fig3}. In Fig.~\ref{fig:fig4}, one sees that for $n_L=1$, the zero-bias conductance is bounded by one conductance quantum. This limitation is imposed by the probe, which offers only one conducting channel. For $n_L>1$, the limit comes from the two  outgoing channels of the sample, and the conductance has an upper limit of $2e^2/h$ irrespective of $n_L$. Finally, at large $t$, $\sigma(0)$ drops as $8(e^2/h)|t_Lt_R|/t^2$ irrespective of the contact size $n_L$.

\begin{figure}[t]
\includegraphics[width=0.9\columnwidth]{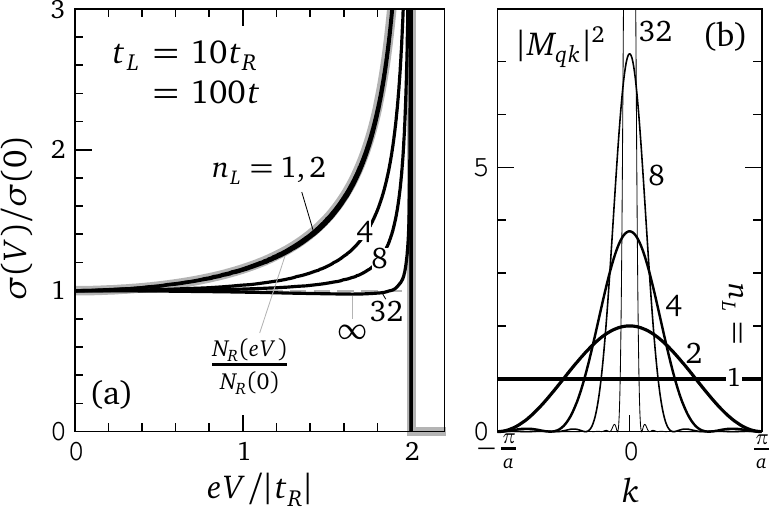}
\caption{\label{fig:fig5}
(a) Differential conductance of the T-shaped junction depicted in Fig.~\ref{fig:fig2} for $t_L=10t_R=100t$ and zero temperature (black lines), compared with the sample LDOS [thick gray line, Eq.~(\ref{eq:T-NR})]. The dashed gray curve for $n_L=\infty$ represents Eq.~(\ref{eq:T-SVinf}) and does not contain the small correction $\delta\sigma(V)$. (b) Function $|M_{qk}|^2$ defined in Eq.~(\ref{eq:Mqk}) for the lowest discrete momentum $q=\pi/(n_L+1)$. The function is unity for $n_L=1$, and approaches $2\pi\delta(k-q)$ as $n_L\to\infty$, imposing the conservation of momentum at the planar junction.
}
\end{figure}

In Fig.~\ref{fig:fig5}(a), we compare the bias-dependence of the differential conductance $\sigma(V)$ with the sample LDOS $N_R(eV)$ in the tunneling regime $t\ll t_L,t_R$. For $n_L=1$ and $n_L=2$, the two quantities are almost undistinguishable, but for $n_L>2$, they start to deviate. Most noticeably, the signature of the band-edge singularity in $N_R(eV)$ is suppressed from the conductance, and the contact becomes ohmic as we approach the planar junction limit $n_L=\infty$. With our choice of parameters, the correction $\delta\sigma$ to the conductance remains small compared to $\tilde{\sigma}$ in the range of biases considered. Therefore, we shall neglect $\delta\sigma$ for the analysis of the results in Fig.~\ref{fig:fig5}. At $n_L=1$, Eq.~(\ref{eq:point_contact}) applies and we can deduce at zero temperature:
	\begin{equation}\label{eq:T-SV1}
		\tilde{\sigma}^{(n_L=1)}(V)=\frac{e^2}{h}
		\frac{4\pi t^2}{|t_L|}\frac{N_R(eV)}{[1+\pi t^2/|t_L|\,N_R(eV)]^2}.
	\end{equation}
This very simple expression results because the real part of the in-chain local propagator $G_R^+(0,0,\omega)$ vanishes within the band, so that only the imaginary part, namely the DOS, remains in the denominator. Equation~(\ref{eq:T-SV1}) confirms that in the point-contact limit, the differential conductance is proportional to the  sample LDOS for $t\to 0$. According to Eq.~(\ref{eq:T-SV1}), the conductance should vanish for $V>2|t_R|$, while the actual conductance turns slightly \emph{negative} above $2|t_R|$ (not shown in the figures). The negative $\delta\sigma(V)$ has a simple interpretation: when the bias reaches the band edge at $2|t_R|$, the current starts to decrease due to the downward curvature of the probe LDOS (see Fig.~\ref{fig:fig3}), resulting in negative differential conductance. This effect cannot be captured by $\tilde{\sigma}(V)$, which neglects the bias dependence of the Green's functions in the probe.

\begin{figure}[t]
\includegraphics[width=0.9\columnwidth]{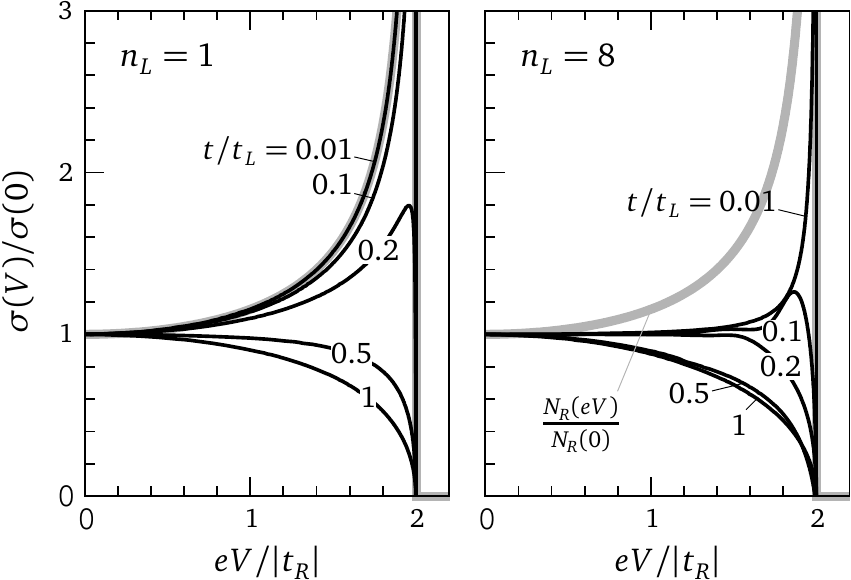}
\caption{\label{fig:fig6}
Differential conductance of the T-shaped junction depicted in Fig.~\ref{fig:fig2} with $n_L=1$ (left panel) and $n_L=8$ (right panel) for $t_L=10t_R$ and various $t$. The thick gray lines show the sample LDOS Eq.~(\ref{eq:T-NR}).
}
\end{figure}

For arbitrary $n_L$, the conductance can be evaluated explicitly (see Appendix~\ref{sec:app3}) as
	\begin{equation}\label{eq:T-SVn}
		\tilde{\sigma}(V)=\frac{e^2}{h}\frac{4\pi t^2}{|t_L|}N_R(eV)\sum_q|\sin q|\,|M_{qk}|^2
		+\mathcal{O}(t^4),
	\end{equation}
where
	\begin{equation}\label{eq:Mqk}
		|M_{qk}|^2=
		\frac{\sin^2q\left\{1-\cos\left[q(n_L+1)\right]
		\cos\left[k(n_L+1)\right]\right\}}{(n_L+1)(\cos q-\cos k)^2}
	\end{equation}
and $\cos k=eV/(2t_R)$. The function $|M_{qk}|^2$ deals with the matching of momenta at the junction. It is displayed in Fig.~\ref{fig:fig5}(b) as a function of $k$ for the lowest $q$ value. At $n_L=1$, there is only one $q$ value, $q=\pi/2$, and $|M_{qk}|^2=1$ so that we recover Eq.~(\ref{eq:T-SV1}). For $n_L=\infty$, we have $|M_{qk}|^2=2\pi\delta(k-q)$ expressing the conservation of momentum at the planar junction. In this limit, the $q$ sum in Eq.~(\ref{eq:T-SVn}) can be converted to an integral and yields $n_L|\sin k|$. Since $k$ is fixed by $V$, we furthermore have $n_L|\sin k|=n_L/[2\pi|t_R|N_R(eV)]$ for $|eV/(2_R)|<1$ and $n_L|\sin k|=0$ otherwise. Hence the LDOS $N_R(eV)$ in Eq.~(\ref{eq:T-SVn}) is exactly canceled in the planar junction and we get ohmic behavior:
	\begin{equation}\label{eq:T-SVinf}
		\tilde{\sigma}^{(n_L=\infty)}(V)=\frac{e^2}{h}n_L\frac{2t^2}{|t_Lt_R|}\theta(2|t_R|-|eV|)
		+\mathcal{O}(t^4).
	\end{equation}
This result can also be derived directly from Eq.~(\ref{eq:local_planar_junction}). It shows that the cancellation of the LDOS singularity in the planar junction requires the exact conservation of momentum at the interface. Although the ohmic conductance of Eq.~(\ref{eq:T-SVinf}) seems to be an instance of Harrison's cancellation argument \cite{Harrison-1961}, it is not. In Harrison's treatment of planar junctions between simple metals, there is a cancellation between the group-velocity and the DOS \emph{in the direction of the current}. The DOS factors in the direction perpendicular to the current are also canceled, but for a different reason, namely, the exponential suppression of large perpendicular momenta in the tunneling matrix element. In our case, the matrix element is momentum independent, and furthermore the sample has no extension along the $x$ direction, so that both ingredients of Harrison's argument are irrelevant.

To finish this section, we point out that, in addition to the kinematic constraint due to conservation of in-plane momentum, the increase of the tunneling amplitude also leads to a suppression of the LDOS singularity in the conductance, as illustrated in Fig.~\ref{fig:fig6}. Even in the point-contact limit, the band-edge singularity is suppressed as $t$ increases. This is obvious from Eq.~(\ref{eq:T-SV1}) since $\tilde{\sigma}^{(n_L=1)}(V)\sim 1/N_R(eV)$ at large $t$. We stress, however, that the validity of our treatment becomes questionable as $t$ increases, since the assumption that the potential drop is confined to the barrier region will loose validity at some point. For example, if $t=t_L$, there is actually no tunnel barrier, and it is not clear where the potential drop should take place.

\section{3D junction to a 2D layer}
\label{sec:3D}

\begin{figure}[t]
\includegraphics[width=0.45\columnwidth]{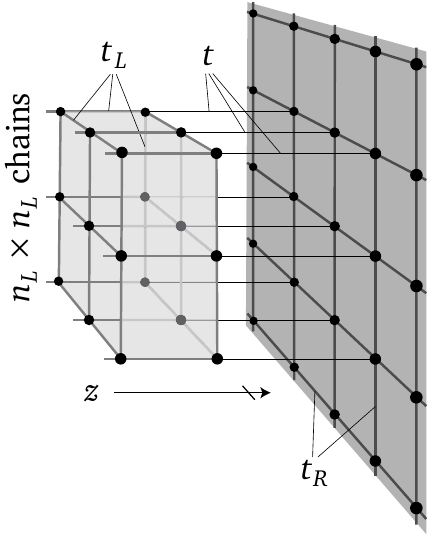}
\caption{\label{fig:fig7}
Tunnel junction to a two-dimensional layer. The probe (left) is semi-infinite in the $z$ direction and has a finite cross section of size $n_L\times n_L$. The sample (right) is an infinite two-dimensional plane. For $n_L=1$ we have a point contact while for $n_L=\infty$ we have a local planar junction following the terminology of Fig.~\ref{fig:fig1}.
}
\end{figure}

In the previous section, we saw that the conservation of in-plane momentum and/or the lowering of the tunnel barrier lead to a suppression of the sample band edge singularity in the tunneling conductance. Here, we investigate the case of a singularity that is not at the band edge, but at zero energy. In the system of Fig.~\ref{fig:fig2}, one could induce a square-root divergence at zero energy by adding next-nearest-neighbor hopping to $\mathcal{H}_R$. The system of Fig.~\ref{fig:fig2} has academic rather than practical interest, however. We consider instead the junction represented in Fig.~\ref{fig:fig7}, which shares some geometrical similarities with realistic systems, in particular the tunnel junctions involving quasi-two-dimensional (2D) layered materials like the high-$T_c$ superconductors. The probe $L$ is a bar of lateral size $n_L\times n_L$ extending from $z=0$ to $z=-\infty$, and the sample $R$ is an infinite plane. Nearest-neighbor sites in $L$ ($R$) are connected by a hopping energy $t_L$ ($t_R$), while neighboring sites at the junction are connected by a tunneling amplitude $t$. As before, we take $L$ and $R$ half-filled for simplicity.

In the sample, the Green's function reads
	\begin{equation}\label{eq:GR-3D}
		G_R^{\pm}(\vec{r}_1,\vec{r}_2,\omega)=\int_{-\pi}^{\pi}\frac{d^2k}{(2\pi)^2}
		\frac{e^{i\vec{k}\cdot(\vec{r}_1-\vec{r}_2)}}{\omega-\varepsilon_{\vec{k}}^R\pm i0^+}
	\end{equation}
with $\varepsilon_{\vec{k}}^R=2t_R(\cos k_x+\cos k_y)$ and $\vec{r}_{1,2}$ the sites of the 2D lattice. The corresponding LDOS $(-1/\pi)\text{Im}\,G_R^+(\vec{r},\vec{r},\omega)$ is translation invariant and presents a logarithmic van Hove singularity at zero energy:
	\begin{equation}\label{eq:3D-NR}
		N_R(\omega)=\frac{1}{2\pi^2|t_R|}K\Big(1-[\omega/(4t_R)]^2\Big)
		\theta(4|t_R|-|\omega|).
	\end{equation}
$K$ is the complete elliptic integral of the first kind. On the $z=0$ surface of the probe, the Green's function for open boundary conditions is a straightforward generalization of Eq.~(\ref{eq:T-GL}):
	\begin{multline}
		G_L^{\pm}(\vec{l}_1,\vec{l}_2,\omega)=\frac{4/t_L}{(n_L+1)^2}\sum_{q_x,q_y}
		\sin(q_xl_{1x})\sin(q_xl_{2x})\\ \times \sin(q_yl_{1y})\sin(q_yl_{2y})
		L\left(\frac{\omega-eV-\varepsilon_{\vec{q}}^L\pm i0^+}{2t_L}\right)
	\end{multline}
with $\varepsilon_{\vec{q}}^L=2t_L(\cos q_x+\cos q_y)$ and $\vec{l}_{1,2}$ the positions of the $n_L^2$ surface sites. The LDOS on the surface presents no divergence, but cusps whose energy scale and amplitude decrease as $n_L$ increases, like those in Fig.~\ref{fig:fig3}. Taking $t_L=10t_R$ is enough to maintain the probe-related conductance correction $\delta\sigma$ small with respect to $\tilde{\sigma}$.

In order to evaluate accurately the Green's function in Eq.~(\ref{eq:GR-3D}), we perform analytically the momentum integration along one direction, and numerically along the other, using fast Fourier transforms with $N=2^{16}$ discrete momenta. The infinitesimal $0^+$ is replaced by $1/N$, which ensures a high energy resolution.

\begin{figure}[t]
\includegraphics[width=0.95\columnwidth]{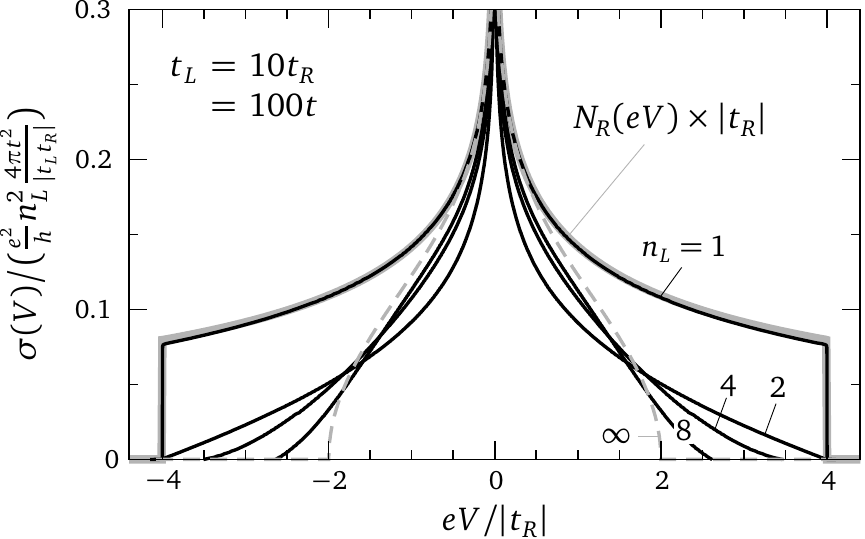}
\caption{\label{fig:fig8}
Differential conductance of the junction depicted in Fig.~\ref{fig:fig7} for $t_L=10t_R=100t$ and zero temperature (black lines), compared with the sample LDOS [thick gray line, Eq.~(\ref{eq:3D-NR})]. The dashed gray curve for $n_L=\infty$ represents Eq.~(\ref{eq:3D-SVinf}) and does not contain the small correction $\delta\sigma(V)$.
}
\end{figure}

\begin{figure}[t]
\includegraphics[width=0.95\columnwidth]{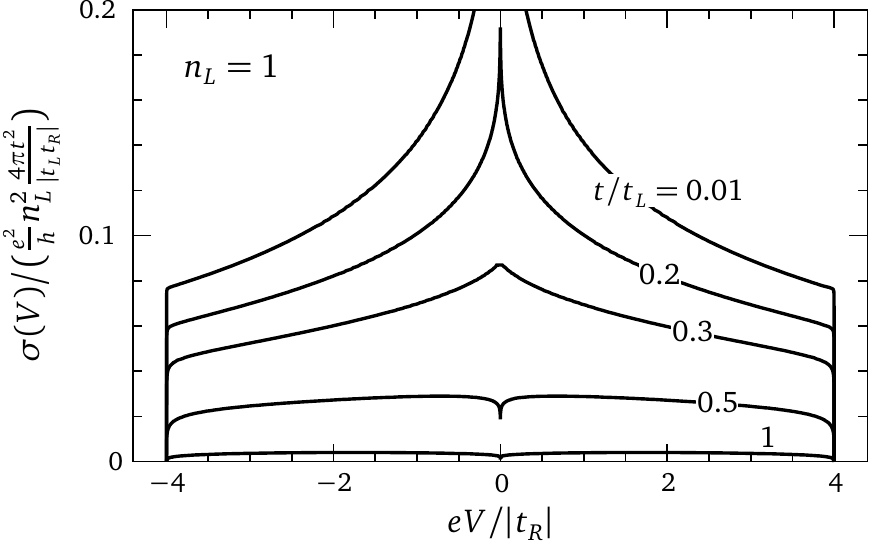}
\caption{\label{fig:fig9}
Zero-temperature differential conductance of the junction depicted in Fig.~\ref{fig:fig7} in the point-contact limit ($n_L=1$) for $t_L=10t_R$ and various values of $t$.
}
\end{figure}

The resulting conductance is displayed in Fig.~\ref{fig:fig8} for $t_L=10t_R=100t$ and various probe sizes $n_L$. At $n_L=1$, the conductance follows accurately the LDOS consistently with Eq.~(\ref{eq:point_contact}), since the junction is a point contact. Something surprising happens at $n_L=2$: the conductance is completely suppressed at the band edges, while the signature of the van Hove singularity remains at the band center. This is due to an interference between two tunneling paths, which is destructive at the band edges. According to the general formula (\ref{eq:current}), there are three types of paths contributing to the current at lowest order in $t$ for $n_L=2$. In the first type, the electron starts from one site on the surface of $L$, tunnels to the sample $R$ and immediately tunnels back to the starting point. This path involves only local propagators in $L$ and $R$. In the second type, the electron tunnels into $R$, propagates to a neighboring site, tunnels back to $L$, and   closes the loop by returning to the initial site. It turns out that this path does not contribute because the advanced and retarded propagators in $L$ are identical at zero energy and cancel [see Eq.~(\ref{eq:sigmatilde})] for nearest-neighbor sites. Finally, in the third path, the electron propagates to a next-nearest-neighbor site in $R$ after tunneling, tunnels back, and closes the loop. This latter path contributes to the current with the same amplitude as the first one, but with the opposite sign because the local and next-nearest neighbor propagators in $L$ are opposite at zero energy. Collecting all terms we find
	\begin{multline}
		\tilde{\sigma}^{(n_L=2)}(V)=-\frac{e^2}{h}\frac{8t^2}{t_L}\text{Im}\!\left[
		G^+_R(0,0,eV)-G^+_R(1,1,eV)\right]\\+\mathcal{O}(t^4)
	\end{multline}
with $G^+_R(0,0,eV)$ and $G^+_R(1,1,eV)$ being the local and next-nearest-neighbor propagators in $R$, respectively. At the band edges, corresponding to $(0,0)$ and $(\pi,\pi)$ states, these two quantities have the same imaginary part, explaining the suppression of the conductance. The behavior of the conductance is qualitatively similar for all sizes $n_L\geqslant 2$: remarkably, the signature of the van Hove singularity remains sharp in the conductance, even in the planar junction limit. This contrasts with the suppression of the square-root singularity in Fig.~\ref{fig:fig5}. For $n_L=\infty$, the conductance can be evaluated analytically starting from Eq.~(\ref{eq:local_planar_junction}); we find
	\begin{multline}\label{eq:3D-SVinf}
		\tilde{\sigma}^{(n_L=\infty)}(V)=\frac{e^2}{h}n_L^2\frac{4\pi t^2}{|t_L|}N_R(eV)\text{Re}
		\sqrt{1-\left(\frac{eV}{2t_R}\right)^2}\\+\mathcal{O}(t^4).
	\end{multline}
Equation~(\ref{eq:3D-SVinf}) is plotted as the dashed gray line in Fig.~\ref{fig:fig8}: in the planar junction the band edges are cut, but the van Hove singularity remains. This behavior is reversed if instead of increasing $n_L$ we keep $n_L=1$ and increase $t$ (see Fig.~\ref{fig:fig9}). In this situation, the conductance remains large close to the band edges, while the van Hove singularity gets suppressed. As discussed at the end of the previous section, however, the validity of the model becomes questionable as $t$ increases.

We can conclude that Harrison's result \cite{Harrison-1961} does not apply in the case of Fig.~\ref{fig:fig7} either: extrapolating the results of Ref.~\onlinecite{Harrison-1961}, one would predict an ohmic behavior for the planar junction, while the actual characteristics is strongly nonlinear, and the effective bandwidth deduced from the conductance curve is two times smaller than the true sample bandwidth. Further discussion of these issues is provided in Sec.~\ref{sec:discussion}.

\section{Momentum-dependence of the matrix element}
\label{sec:STM}

\begin{figure}[b]
\includegraphics[width=0.45\columnwidth]{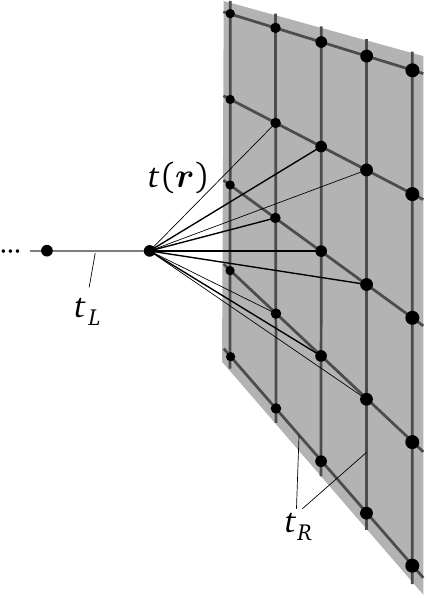}
\caption{\label{fig:fig10}
``STM-like'' junction to a two-dimensional layer. The probe is a semiinfinite nearest-neighbor chain with hopping $t_L$, and the sample is an infinite two-dimensional plane with hopping $t_R$. Tunneling can take place from the extremity of the probe to various sites in the sample.
}
\end{figure}

In this section, we discuss the most prominent effects associated with a momentum dependence of the tunneling matrix element. For this purpose, we consider the system represented in Fig.~\ref{fig:fig10}. The sample is an infinite two-dimensional layer like in the preceding section. It's Green's function and local DOS are given by Eqs.~(\ref{eq:GR-3D}) and (\ref{eq:3D-NR}), respectively. The probe is a semiinfinite chain corresponding to the $n_L=1$ limit in Secs.~\ref{sec:T} and \ref{sec:3D}. The tunneling is not constrained to be local, but can take place from the extremity of $L$ to various sites $\vec{r}$ in $R$ with an amplitude $t(\vec{r})$. Figure~\ref{fig:fig10} is a particular case of the ``STM-like'' junction sketched in Fig.~\ref{fig:fig1}. We will focus on describing how the matrix element $t(\vec{r})$ can change the simple proportionality, which is realized in the local limit, between the conductance and the LDOS.

The main contribution to the conductance at zero temperature can be deduced from Eq.~(\ref{eq:STM_junction}). Making use of the translational invariance in the sample we find
	\begin{equation}
		\tilde{\sigma}(V)=\frac{e^2}{h}\frac{4\pi}{|t_L|}\frac{\frac{1}{N}\sum_{\vec{k}}
		|t(\vec{k})|^2A_R(\vec{k},eV)}{\left|1+\frac{i}{|t_L|}
		\frac{1}{N}\sum_{\vec{k}}|t(\vec{k})|^2G_R^+(\vec{k},eV)\right|^2},
	\end{equation}
where $G_R^+(\vec{k},eV)=(eV-\varepsilon_{\vec{k}}^R+i0^+)^{-1}$ with $\varepsilon_{\vec{k}}^R=2t_R(\cos k_x+\cos k_y)$, $A_R(\vec{k},eV)=\delta(eV-\varepsilon_{\vec{k}}^R)$ is the spectral function, $t(\vec{k})$ is the Fourier transform of $t(\vec{r})$, and $N$ is the number of $\vec{k}$ points. We shall restrict to the tunneling regime $t(\vec{r})\ll t_L,t_R$, in which case we have the simple result
	\begin{equation}\label{eq:STM}
		\tilde{\sigma}(V)=\frac{e^2}{h}\frac{4\pi}{|t_L|}\frac{1}{N}\sum_{\vec{k}}
		|t(\vec{k})|^2A_R(\vec{k},eV)+\mathcal{O}(t^4).
	\end{equation}
Hence the conductance is an average over sample states in reciprocal space, each state being weighted by $|t(\vec{k})|^2$. A formula analogous to Eq.~(\ref{eq:STM}) has been occasionally used to analyze tunneling spectra of the cuprates.\footnote{See Ref.~\onlinecite{Fischer-2007}, and references therein.} We stress here, though, that the validity of Eq.~(\ref{eq:STM}) does not extend beyond the simple model of Fig.~\ref{fig:fig10}. In particular, this expression cannot capture interference effects taking place within the probe, such as those responsible for the suppression of the band-edge conductance in Fig.~\ref{fig:fig8}.

\begin{figure}[tb]
\includegraphics[width=\columnwidth]{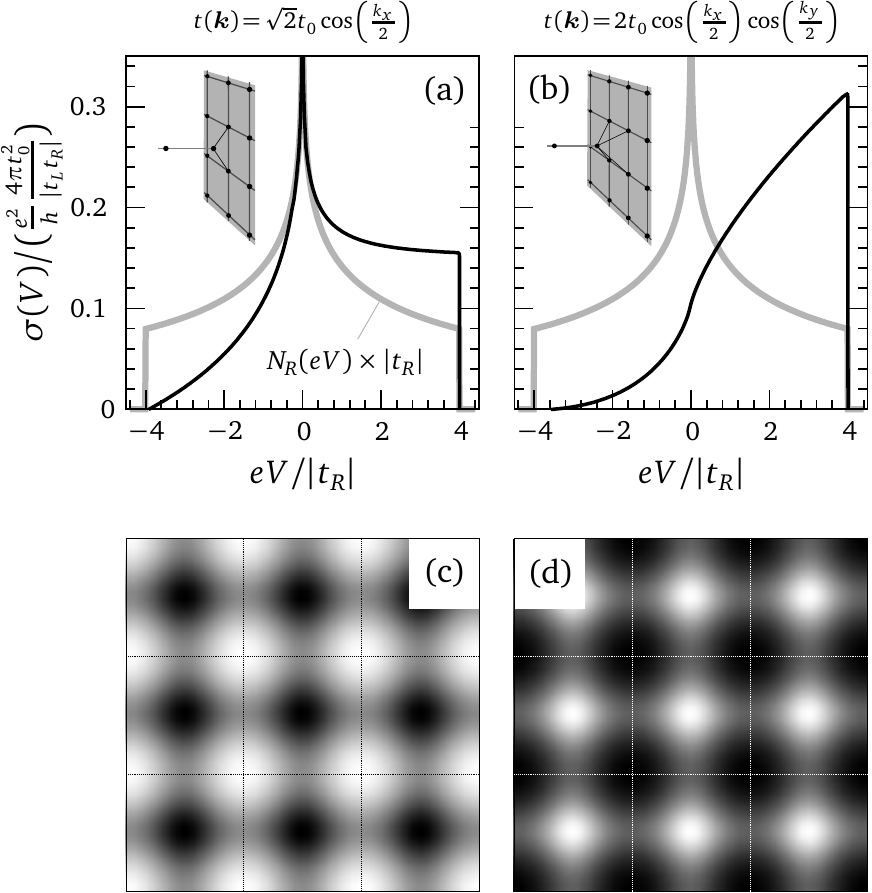}
\caption{\label{fig:fig11}
(a) Zero-temperature differential conductance of the junction depicted in Fig.~\ref{fig:fig10}, with a tunneling amplitude given by $t_0/\sqrt{2}$ for one pair of neighboring sites, and zero otherwise (see the inset). The momentum dependence of the matrix element is indicated above the graph; the sample LDOS is displayed in gray for comparison. (b) Same as (a) with a tunneling amplitude given by $t_0/2$ for four sites around a plaquette, and zero otherwise. (c) Map of the tip-sample distance obtained in constant-current mode using the matrix element (\ref{eq:tr}) with $\kappa=5$. The distance is largest (white) above the lattice sites. (d) Corresponding conductance map evaluated at $eV=2t_R$. The conductance is largest (white) between the lattice sites. The model parameters are $t_L=10t_R=100t_0$.
}
\end{figure}

The simplest situation is the local limit $t(\vec{r})=t_0\delta_{\vec{r}\vec{r}_0}$, or $t(\vec{k})=t_0$. In this case, the conductance reflects accurately the DOS as we have seen in the previous section; this is also clear from Eq.~(\ref{eq:STM}) since $(1/N)\sum_{\vec{k}}A_R(\vec{k},eV)=N_R(eV)$. The second case of interest is when the ``tip'' is located in-between two sites of the sample, above a nearest-neighbor bond. To begin with, it is most natural to assume that the tunneling amplitude is the same for the two nearest-neighbor sites below the tip, and zero otherwise. The resulting differential conductance is shown in Fig.~\ref{fig:fig11}(a), and deviates significantly from the DOS: it is suppressed at the lower band edge and enhanced at the upper edge. This is easily understood considering that the matrix element $t(\vec{k})\sim\cos(k_x/2)$ or $\cos(k_y/2)$ depending on the bond orientation. In either case, the matrix element vanishes at $(\pi,\pi)$ (lower band edge) and is largest at $\Gamma=(0,0)$ (upper band edge). The asymmetric conductance can also be interpreted as the interference between two types of tunneling paths in real space. The first path involves the local propagators in $L$ and $R$, while the second involves the local propagator in $L$ and the nearest-neighbor propagator in $R$; we thus find $\tilde{\sigma}(V)\propto-\text{Im}\!\big[G^+_R(0,0,eV)+G^+_R(1,0,eV)\big]$ at lowest order in $t_0$. For a $(\pi,\pi)$ state $G^+_R(0,0,eV)$ and $G^+_R(1,0,eV)$ cancel, while for a $\Gamma$ state they add up.

Let's now assume that the tip is located above the center of a plaquette. The tunneling amplitudes are equal for the four sample sites forming the plaquette, and they vanish elsewhere. The resulting conductance displayed in Fig.~\ref{fig:fig11}(b) is strongly asymmetric, this time the signature of the van Hove singularity is completely suppressed, owing to the matrix element being zero both at $(\pi,\pi)$ and $(\pi,0)$. Considering the various tunneling paths, we get in this case $\tilde{\sigma}(V)\propto-\text{Im}\!\big[G^+_R(0,0,eV)+2G^+_R(1,0,eV)+G^+_R(1,1,eV)\big]$.

The two extreme cases of Figs.~\ref{fig:fig11}(a) and \ref{fig:fig11}(b) suggest that by scanning a tip over a discrete lattice, one can observe large variations of the local tunneling conductance if the ``tunneling cone'' is sufficiently narrow, i.e., if only the sites closest to the tip contribute significantly to the current. As a generalization of these two extreme cases, we now consider a tunneling amplitude that is a continuous function of the tip position $\vec{l}$, decaying exponentially with the distance to the tip:
	\begin{equation}\label{eq:tr}
		t(\vec{r})=t_0e^{-\kappa(|\vec{r}-\vec{l}|-d_0)}.
	\end{equation}
$d_0$ is a reference distance set equal to the sample lattice parameter. $\kappa$ controls the opening of the tunneling cone. We may relate $\kappa$ to an effective work function $\phi$ of the junction through the relation $\phi/E_{\text{F}}=(\kappa/k_{\text{F}})^2\sim(\kappa/\pi)^2$. $\phi$ thus varies between 0.4 and 2.5 times $E_{\text{F}}$ for $\kappa$ ranging between two and five in units of the sample lattice. The vertical tip-sample separation $d$ (out-of-plane coordinate of $\vec{l}$) is adjusted in order to keep the total current fixed, in the fashion of a constant-current STM experiment. The reference current is set to the value calculated at $eV=2t_R$ when the tip sits on top of a lattice site. For $\kappa=2$, the tunneling cone is wide and the current averages over several paths. As a result, the conductance is almost uniform when scanning the tip over the lattice. Furthermore, it is strongly asymmetric, almost completely suppressed at negative bias and peaked at positive bias. Indeed, as $\kappa\to0$, we approach the limit $t(\vec{k})=t_0\delta_{\vec{k},\vec{0}}$ where tunneling is prohibited except for the $\Gamma$ state at the upper band edge. The situation is more interesting for $\kappa=5$, where the tunneling cone is narrower and the matrix element interpolates smoothly between the situations shown in Figs.~\ref{fig:fig11}(a) and \ref{fig:fig11}(b). In order to maintain the current when moving away from a lattice site, the tip has to approach the sample. The resulting height field displayed in Fig.~\ref{fig:fig11}(c) has a corrugation $\Delta d/d\approx14\%$. Figure~\ref{fig:fig11}(d) shows a conductance map calculated simultaneously at $eV=2t_R$. Remarkably, the regions near the plaquette centers appear with the largest conductance, because the total current is fixed, and at these locations the conductance increases with increasing bias [Fig.~\ref{fig:fig11}(b)]. This can be understood as follows: assume that the conductance varies as $\sigma(V)\propto V^{\alpha}$ with a position-dependent exponent $\alpha$, then the constant-current condition $I(V_0)=I_0$ implies that $\sigma(V_0)\propto 1+\alpha$. Therefore it appears that in constant-current mode, a conductance map gives an image of the bias dependence of the local conductance, rather than a horizontal cut of the LDOS. A map of the conductance at fixed tip height, in contrast, provides an image similar to Fig.~\ref{fig:fig11}(c).

\section{Discussion}
\label{sec:discussion}

The two models studied in Secs.~\ref{sec:T} and \ref{sec:3D} show that there can be significant qualitative differences between the conductances measured at planar and point-contact junctions to the same given sample. The conservation of in-plane momentum at the planar junction leads in one case to a suppression of the band-edge singularity (Sec.~\ref{sec:T}), in the other case to the complete suppression of the conductance at the band edge without affecting the singularity at the band-center (Sec.~\ref{sec:3D}). One can conclude that the differential conductance is not in general proportional to the sample DOS at planar junctions, although features of the DOS may be present in the conductance, like e.g., the van Hove singularity in Fig.~\ref{fig:fig8}. There seems to be no simple rule to determine \textit{a priori} whether a singularity of the DOS will or will not show up in the conductance of a planar junction. In contrast, at the point contact, a simple proportionality is found between the conductance and the sample LDOS, if the tunneling amplitude is small. When interpreting tunneling data measured at real tunnel junctions, one has to wonder which one of these two limits is the best theoretical standpoint. Both planar and point-contact limits are appealing theoretically, as they generally do not require a microscopic modeling of the junction. They are not equivalent, though. The use of planar-junction models \cite{Harrison-1961, Duke-1969, Wolf-1985, Ledvij-1995} for interpreting STM data may therefore appear questionable, since the STM junction is clearly closer to the point contact than to the planar-junction limit \cite{Chen-1988, Wiesendanger-1993}. Planar-junction models may not even be appropriate for break junctions, in which the uncontrolled roughness of the interface can prevent conservation of in-plane momentum. The similitude of the tip-sample junction with a point-contact is supported by first-principle investigations, indicating that the STM is most likely a microscopic atom-to-atom contact \cite{Tsukada-1993}. Once formulated in terms of localized orbitals (e.g., Wannier functions), the tunneling problem may reduce in this case to a point contact involving only two localized orbitals, and thereby lead to a tunneling conductance given by Eq.~(\ref{eq:point_contact}), i.e., proportional to the sample LDOS in the appropriate limits.

The applications considered in the present study are simple noninteracting single-band tight-binding models, but the formalism we have used has a broader validity. For instance, Eqs.~(\ref{eq:current})--(\ref{eq:planar_junction}) remain unchanged if the electrodes are multiband systems with interactions---the additional complexity being embodied in the real-space Green's functions---provided that there is no interaction between the electrodes. It is also possible to extend the formalism to continuum  models; here, the main difficulty lies in the definition of the tunneling amplitude in Eq.~(\ref{eq:tunneling-Hamiltonian}) for continuous spatial coordinates. The case of superconducting electrodes requires further work. Due to the nonconservation of the particle number in the electrodes, additional terms involving anomalous propagators appear in the expression of the current, corresponding to Andreev and Josephson contributions. This problem has been investigated for simple geometries using the tunneling-Hamiltonian method and the Keldysh nonequilibrium formalism \cite{Cuevas-1996, Yan-2000, Zeng-2003, Bolech-2004, *Bolech-2005, Bezuglyi-2006, *Bezuglyi-2007}, but the role of the junction' geometry for tunneling contacts involving superconductors remains largely unexplored.

Nevertheless, thanks to its sub-meV resolution, tunneling is particularly useful for investigating the excitation spectrum of superconductors, whose pairing gap can be small---typically 1~meV for a $T_c$ of 7~K. This was beautifully illustrated by Giaever and his demonstration of the superconducting gap \cite{Giaever-1960}, and later by the detailed observation of vortex cores \cite{Hess-1989, *Hess-1990, *Hess-1990b, Renner-1991, Maggio-Aprile-1995, Hoogenboom-2000a, *Hoogenboom-2000b, *Hoogenboom-2001, Pan-2000b, Hoffman-2002, Eskildsen-2002, Matsuba-2003a, *Matsuba-2007, Nishimori-2004, Levy-2005, Yin-2009, Teague-2009, *Beyer-2009} and impurity resonances \cite{Yazdani-1997, *Yazdani-1999, Hudson-1999, *Pan-2000a, *Hudson-2001, *Hudson-2003, Chatterjee-2008, Machida-2010} by STM. The early theory \cite{Schrieffer-1963} has lead to the idea that the differential conductance is proportional to an effective DOS carrying information about the superconducting gap function, but no information about the band structure. Looking closer, it turns out that the information on the underlying band structure was discarded from the start, rather than proven to be absent from the tunneling spectrum. This theory also neglects any momentum dependence of the matrix element. On the contrary, the theory of Tersoff and Hamann \cite{Tersoff-1983} for the STM is based on an explicit calculation of the matrix element, and leads to the idea that the conductance measures the sample LDOS at the position of the tip. In this scenario, the experimental STM tunneling conductance should be compared with the full LDOS containing all band-structure and many-body effects, as was done in recent years for optimally-doped cuprate superconductors of the bismuth family \cite{Hoogenboom-2003b, Levy-2008, Jenkins-2009}. It was pointed out in Ref.~\onlinecite{Berthod-2010} that the analysis of STM data for quasi-2D cuprates by means of the effective DOS concept can lead to erroneous conclusions.

In the Bi-based cuprates, the STM spectra develop with underdoping a more and more asymmetric shape, with excess weight at negative bias (occupied states) \cite{Fischer-2007}. This has been attributed to the onset of correlations when approaching the Mott insulating phase \cite{Rantner-2000, Randeria-2005, Anderson-2006}. Another interpretation, based on atomistic modeling of the tip-sample system, was recently proposed, in terms of copper $d_{z^2}$ states and material-specific selection rules \cite{Nieminen-2009}. The asymmetric conductances in Fig.~\ref{fig:fig11} suggest yet another interpretation, based on the momentum dependence (rather than the energy dependence) of the tunneling matrix element, to explain at least part of the asymmetry in the cuprate STM spectra. A matrix element of the form (\ref{eq:tr}) invariably favors the zone-center states, and leads for the model of Fig.~\ref{fig:fig10} to an asymmetric conductance with excess weight at positive bias (zone-center states have positive energy in Sec.~\ref{sec:STM} because $t_R>0$). In the cuprates, the zone-center states lie at negative energy, so that this same mechanism would produce an asymmetry with excess weight at negative bias, as observed experimentally. In addition to overweighting the zone-center states, the matrix element of Eq.~(\ref{eq:tr}) introduces local variations of this weighting, yielding more asymmetric conductance spectra in between lattice sites. If the matrix element indeed contributes to the conductance asymmetry in STM junctions to the cuprates, one expects to see variations of this asymmetry when moving the tip over subunit cell distances. Interestingly, recent investigations of the conductance asymmetry have unveiled such subnanometer variations, with enhanced asymmetry at the center of the plaquette formed by copper sites \cite{Kohsaka-2007, *Lawler-2010}.

\section{Conclusion}

We have explored the relationship between the differential tunneling conductance and the local density of states (LDOS) in tunnel junctions, in particular the role of the junction' geometry. Our approach has been to solve exactly within the Keldysh formalism simple tight-binding models in which the geometry can be tuned from the limit of a point contact to the limit of a planar contact. While the conductance is simply proportional to the sample LDOS at tunneling point contacts, there can be significant differences between the conductance and the sample LDOS at planar junctions, including the complete absence in the conductance of divergences present in the LDOS, which can be suppressed by kinematic effects (in-plane momentum conservation) or by interference between several tunneling paths. We have also studied a simple model of STM junction, and shown how the dispersion of electrons from the STM tip can induce strong variations of the local $I(V)$ characteristics, and generate a contrast in the constant-current topographic image that is not related to variations of the LDOS.

\acknowledgments

We are grateful to {\O}.\ Fischer and M.\ B{\"u}ttiker for stimulating discussions and to A.-P.\ Jauho for his comments. This work was supported by the Swiss National Science Foundation through Division II and MaNEP.

\appendix

\section{Calculation of the current}
\label{sec:app1}

The tunnel current can be evaluated in the usual way \cite{Mahan-2000}, as the rate of change of the particle number in $R$:
	\begin{equation}
		I=e\langle\dot{N}_R\rangle
	\end{equation}
with $N_R=\sum_{\vec{r}}\psi^{\dagger}(\vec{r})\psi(\vec{r})$. Our convention is $e=|e|$, so that the current is positive when electrons flow from $L$ to $R$, a positive bias $V$ being applied to $R$. This corresponds to the usual convention adopted in tunneling spectroscopy. Since only $\mathcal{H}_T$ can change $N_R$, we have
	\begin{eqnarray}
		\nonumber
		i\hbar\dot{N}_R&=&[N_R,\mathcal{H}]=[N_R,\mathcal{H}_T]\\
		&=&-\sum_{\vec{l}\vec{r}}T(\vec{l},\vec{r})\left[\psi^{\dagger}(\vec{l})\psi(\vec{r})-
		\psi^{\dagger}(\vec{r})\psi(\vec{l})\right]
	\end{eqnarray}
and the current becomes
	\begin{eqnarray}\label{eq:Acurrent1}
		\nonumber
		I&=&\frac{ie}{\hbar}\sum_{\vec{l}\vec{r}}T(\vec{l},\vec{r})\left[
		\langle\psi^{\dagger}(\vec{l})\psi(\vec{r})\rangle-
		\langle\psi^{\dagger}(\vec{r})\psi(\vec{l})\rangle\right]\\
		&=&\frac{-2e}{\hbar}\sum_{\vec{l}\vec{r}}T(\vec{l},\vec{r})\text{Im}\,
		\langle\psi^{\dagger}(\vec{l})\psi(\vec{r})\rangle.
	\end{eqnarray}
The quantity $\langle\psi^{\dagger}(\vec{l})\psi(\vec{r})\rangle$ corresponds to the lesser Green's function at time $t=0^+$ \cite{Rammer-1986}. It can be expressed in terms of the usual retarded Green's function $G^+(\vec{r},\vec{l},t)=-i\theta(t)\langle[\psi(\vec{r},t),\psi^{\dagger}(\vec{l},0)]_+\rangle$ and the Keldysh function $G^K(\vec{r},\vec{l},t)=-i\langle[\psi(\vec{r},t),\psi^{\dagger}(\vec{l},0)]\rangle$:
	\begin{equation}
		\langle\psi^{\dagger}(\vec{l})\psi(\vec{r})\rangle=\frac{i}{2}
		\left[G^+(\vec{r},\vec{l},0^+)-G^K(\vec{r},\vec{l},0^+)\right].
	\end{equation}
Introducing this into Eq.~(\ref{eq:Acurrent1}), we see that, considering the quantities $T(\vec{l},\vec{r})$ and $G(\vec{r},\vec{l},0^+)$ as matrices in the $\{\vec{l},\vec{r}\}$ space, the $\vec{r}$ sum is equivalent to a matrix product and the $\vec{l}$ sum to a trace. Noting further that $G(0^+)=\int\frac{d\omega}{2\pi}\,G(\omega)$, we obtain
	\begin{equation}
		I=\frac{e}{h}\int d\omega\,\text{Re}\,\text{Tr}\,T[G^K(\omega)-G^+(\omega)].
	\end{equation}
On general grounds, we have that $\int d\omega\,\text{Re}\,G^+(\omega)=0$. This can be deduced from the representation of the retarded function in terms of the spectral function $\rho(\varepsilon)$:
	\begin{equation}
		\text{Re}\,G^+(\omega)=\int\hspace{-1.2em}\mathscr{P}\,
		d\varepsilon\,\frac{\rho(\varepsilon)}{\omega-\varepsilon}.
	\end{equation}
Hence we find Eq.~(\ref{eq:current}) for the current.

In order to evaluate $G^K(\omega)$ in Eq.~(\ref{eq:current}), we solve the Dyson's equation in Keldysh space, using the same representation as in Ref.~\onlinecite{Rammer-1986}:
	\begin{equation}
		\mathbb{G}=\begin{pmatrix}G^+&G^K\\0&G^-\end{pmatrix},
	\end{equation}
with $G^-$ the usual advanced function. The Dyson equation for $\mathbb{G}\equiv\mathbb{G}(\vec{r},\vec{l},\omega)$ expresses all ways of propagating an electron from $\vec{r}$ in $R$ to $\vec{l}$ in $L$ and reads
	\begin{equation}\label{eqA:Dyson}
		\mathbb{G}=\mathbb{G}_R\mathbb{T}\mathbb{G}_L+\mathbb{G}_R\mathbb{T}\mathbb{G}_L\mathbb{T}\mathbb{G}_R
		\mathbb{T}\mathbb{G}_L+\ldots
	\end{equation}
We have introduced the propagators of the isolated electrodes, $\mathbb{G}_L$ and $\mathbb{G}_R$, as well as a matrix representing the tunneling term, which is diagonal in the Keldysh indices,
	\begin{equation}
		\mathbb{T}=\begin{pmatrix}T&0\\0&T\end{pmatrix}.
	\end{equation}
Equation~(\ref{eqA:Dyson}) implies matrix products in both the Keldysh and site indices. The term of first order in $\mathbb{T}$, for instance, should be understood as
	\[
		\sum_{\vec{r}_1\vec{l}_1}\mathbb{G}_R(\vec{r},\vec{r}_1,\omega)
		\mathbb{T}(\vec{r}_1,\vec{l}_1)\mathbb{G}_L(\vec{l}_1,\vec{l},\omega),
	\]
since the propagators $\mathbb{G}_R$ and $\mathbb{G}_L$ are restricted to the subsystems $R$ and $L$, respectively. Summing the infinite series of terms in Eq.~(\ref{eqA:Dyson}) leads to
	\begin{multline}\label{eqA:Dyson2}
		\mathbb{G}=\mathbb{G}_R\mathbb{T}\mathbb{G}_L(\openone-\mathbb{T}\mathbb{G}_R\mathbb{T}\mathbb{G}_L)^{-1}\\
		=\left[(\mathbb{G}_R\mathbb{T}\mathbb{G}_L)^{-1}-\mathbb{T}\right]^{-1}.
	\end{multline}
The matrix inversions in the Keldysh indices can be done explicitly, yielding the following expression for the Keldysh component $G^K$:
	\begin{multline}
		G^K=(\openone-G_R^+TG_L^+T)^{-1}(G_R^+TG_L^K+G_R^KTG_L^-)\\
		\times(\openone-TG_R^-TG_L^-)^{-1}.
	\end{multline}
At this point, we use the assumption that each electrode is in thermal equilibrium and is characterized by its own chemical potential, say $\mu_L=\mu_R+eV$ and $\mu_R$. The potential drop $eV$ takes place in the barrier and does not affect $L$ and $R$. Hence $G_{L,R}^{\pm,K}$ should be evaluated for the isolated $L$ and $R$ systems at thermal equilibrium. In conditions of thermal equilibrium, all Green's functions can be expressed in terms of only two spectral functions: $G^K$ is therefore not independent of $G^+$ and $G^-$. Specifically, we have the property $G_{L,R}^K(\omega)=[1-2f_{L,R}(\omega)][G_{L,R}^+(\omega)-G_{L,R}^-(\omega)]$ with $f_{L,R}$ the Fermi distribution measured with respect to $\mu_{L,R}$, and we thus obtain Eq.~(\ref{eq:GK}).

\section{Equivalence with Todorov \textit{et al.}}
\label{sec:app2}

Here we show that Eqs.~(\ref{eq:current}) and (\ref{eq:GK}) give the same current as the formula obtained by Todorov \textit{et al.} in Ref.~\onlinecite{Todorov-1993}. These authors expressed the current in terms of the t-matrix, which in our notation reads
	\begin{multline}\label{eq:t-matrix}
		t=T+TG_L^+TG_R^+T+TG_L^+TG_R^+TG_L^+TG_R^+T+\ldots\\
		=T(\openone-G_L^+TG_R^+T)^{-1}.
	\end{multline}
We first note the identity
	\begin{multline}\label{eq:TrRe}
		\text{Tr}\,\text{Re}\,(\openone-TG_R^+TG_L^+)^{-1}TG_R^+T(G_L^+-G_L^-)
		(\openone-TG_R^-TG_L^-)^{-1}=\\
		-\text{Tr}\,\text{Re}\,(\openone-TG_R^+TG_L^+)^{-1}T(G_R^+-G_R^-)TG_L^-
		(\openone-TG_R^-TG_L^-)^{-1}.
	\end{multline}
This can be proven by taking the conjugate to develop the real part, and then by checking order by order in $T$. The relation (\ref{eq:TrRe}), together with the cyclic property of the trace, allow to rewrite the current Eqs.~(\ref{eq:current}) and (\ref{eq:GK}) as
	\begin{multline}
		I=\frac{e}{h}\int d\omega\,\text{Tr}\,\text{Re}\,(G_L^--G_L^+)(\openone-TG_R^-TG_L^-)^{-1}\\
		\times(\openone-TG_R^+TG_L^+)^{-1}TG_R^+T(2f_L-2f_R).
	\end{multline}
Conjugating to evaluate the real part, and using the relations $(\openone-TG_R^+TG_L^+)^{-1}TG_R^+T=TG_R^+T(\openone-G_L^+TG_R^+T)^{-1}$ and $TG_R^-T(\openone-G_L^-TG_R^-T)^{-1}=(\openone-TG_R^-TG_L^-T)^{-1}TG_R^-T$, this can be rearranged as
	\begin{multline}
		I=\frac{e}{h}\int d\omega\,\text{Tr}\,(G_L^--G_L^+)(\openone-TG_R^-TG_L^-)^{-1}T\\
		\times(G_R^+-G_R^-)T(\openone-G_L^+TG_R^+T)^{-1}(f_L-f_R).
	\end{multline}
One recognizes the t-matrix of (\ref{eq:t-matrix}) as well as the spectral functions $G_{L,R}^--G_{L,R}^+=2\pi i\rho_{L,R}$, following the notation of Ref.~\onlinecite{Todorov-1993}, which gives Eq.~(37) of Ref.~\onlinecite{Todorov-1993}:
	\begin{equation}
		I=\frac{2\pi e}{\hbar}\int d\omega\,\text{Tr}\,\rho_Lt^{\dagger}\rho_Rt\,(f_L-f_R).
	\end{equation}

\section{T-shaped junction for weak tunneling}
\label{sec:app3}

For the junction shown in Fig.~\ref{fig:fig2}, at lowest order in $t$ and zero temperature, the conductance $\tilde{\sigma}(V)$ of Eq.~(\ref{eq:sigmatilde}) reduces to
	\begin{multline}
		\tilde{\sigma}(V)=\frac{e^2}{h}2t^2\sum_{l_1=1}^{n_L}\sum_{l_2=1}^{n_L}
		\text{Re}\,G_R^+(l_1-l_2,eV)\\ \times[G_L^-(l_2,l_1,eV)-G_L^+(l_2,l_1,eV)].
	\end{multline}
Using the property $L(x\pm i0^+)=x\mp i\text{Re}\sqrt{1-x^2}$ valid for $|x|<1$, we find from Eq.~(\ref{eq:T-GL}) that
	\begin{multline}
		G_L^-(l_2,l_1,eV)-G_L^+(l_2,l_1,eV)=\\
		\frac{4 i/|t_L|}{n_L+1}\sum_q\sin(ql_2)\sin(ql_1)|\sin q|.
	\end{multline}
Furthermore, we have from Eq.~(\ref{eq:T-GR}) that
	\begin{multline}
		\text{Re}\,iG_R^+(l_1-l_2,eV)=-\text{Im}\,G_R^+(l_1-l_2,eV)=\\
		\pi\int_{-\pi}^{\pi}\frac{dk}{2\pi}\,e^{ik(l_1-l_2)}\delta(eV-2t_R\cos k).
	\end{multline}
Collecting the terms, we find
	\begin{equation}
		\tilde{\sigma}(V)=\frac{e^2}{h}\frac{4\pi t^2}{|t_L|}\sum_q|\sin q|
		\int_{-\pi}^{\pi}\frac{dk}{2\pi}\,|M_{qk}|^2\delta(eV-2t_R\cos k)
	\end{equation}
where we have defined
	\begin{equation}
		|M_{qk}|^2=\frac{2}{n_L+1}\left|\sum_{l=1}^{n_L}\sin(ql)e^{ikl}\right|^2.
	\end{equation}
The expression of $|M_{qk}|^2$ can be worked out analytically for arbitrary $n_L$; the result is displayed in Eq.~(\ref{eq:Mqk}). Finally, the formula (\ref{eq:T-SVn}) results by noting that
	\begin{equation}
		\delta(eV-2t_R\cos k)=2\pi N_R(eV)\delta(k-k_0)
	\end{equation}
with $\cos k_0=eV/(2t_R)$.

\bibliography{}

\end{document}